# Are Unidentified EUV Sources the Closest Neutron Stars?

A. Shemi

*Wise Observatory & School of Physics and Astronomy*
*Raymond and Beverly Sackler Faculty of Exact Sciences, Tel Aviv University, Tel Aviv, 69978, Israel.*

**ABSTRACT**

Unidentified extreme UV sources, detected in the *EUVE* and *ROSAT* WFC all-sky surveys, could be isolated old neutron stars, accreting material from the ISM. The closest neutron stars, that are located in the local ISM bubble of unusually low density, are faint and cool ($L \sim 10^{27}$ erg s$^{-1}$, $T \lesssim 6$eV). The extreme UV spectrum of these sources is very sensitive to the H I column density, since large fraction of the emission is emitted just below the hydrogen Lyman edge. The *EUVE* sources with large count rate in the long-wave bandpass (600Å) seem to be the most promising candidates. These sources should have low H I column density ($N_{HI} \lesssim 10^{18}$ cm$^{-2}$), constraining their distance to a few tens of parsec. Otherwise, their spectra would be significantly modified by ISM absorption, inevitably become stronger in the short wavelength (100Å, 200Å) *EUVE* bandpasses. If these unidentified objects are familiar EUV sources rather than neutron stars, either white dwarfs, late type stars or cataclysmic variables, they are expected to be identifiable, generally brighter than $V \sim 14$.

## 1 INTRODUCTION

About 10% of the sources detected in the all sky surveys of the *EUVE* (Malina et al. 1994, Bowyer et al. 1994) and the *ROSAT* WFC (Pound et al. 1993) have no previous identification at optical or other energies. The nature of these sources, generically classified as 'NOIDs', is unclear and probably inhomogeneous. There are no strong constraints on their distance, some of the NOIDs may be extragalactic and some may contain variable and flaring sources. However, there is the exciting possibility that some of them might be nearby isolated old neutron stars (i.e. neutron stars not in binaries or in supernova remnants, and too old to emit radiation as normal pulsars) accreting material from the ISM (Ostriker, Rees & Silk 1970, Treves & Colpi 1991, Blaes & Madau 1993). Our Galaxy is estimated to have $\sim 10^9$ isolated old neutron stars, a population 3-4 orders of magnitude larger than the pulsar population, but that has so far received little attention. The closest isolated neutron star is likely to be less than $\sim 10$ pc from Earth, and several hundreds are expected within 100 pc.

The *EUVE* survey, lasted from 24 July 1992 to 21 January 1993, covers 97% of the sky with exposures ranging from a few hundred seconds at the ecliptic to 20 kiloseconds at the poles. The survey was carried out in parallel with three scanners, and a Deep Survey/Spectrometer (DS/S) telescope, pointed in perpendicular to the scanners. The peak wavelength and wavelength regions at 10% filter bandpasses are

1) Scanner A & B, filter Lexan/B ("Lex") $\bar{\lambda} = 89$Å $\Delta\lambda = 58 - 174$Å; filter Al/Ti/C ("AlC") $\bar{\lambda} = 171$Å $\Delta\lambda = 156 - 234$Å;

2) Scanner C, filter Al/Ti/C/Sb ("Dagwood") $\bar{\lambda} = 405$Å $\Delta\lambda = 345 - 605$Å; filter Si/SiO ("Tin") $\bar{\lambda} 551$Å, $\Delta\lambda = 519 - 743$Å;

3) DS/S, filter Lexan/B $\bar{\lambda} = 91$Å $\Delta\lambda = 67 - 178$Å; filter Al/C $\bar{\lambda} = 171$Å, $\Delta\lambda = 157 - 364$Å.

The *ROSAT* survey, started on July 1991 and lasted about half a year, was performed with two filters, S1, $\Delta\lambda = 60 - 140$Å, and S2, $\Delta\lambda = 110 - 200$Å.

The unidentified EUV sources detected by *EUVE* and *ROSAT* WFC are generally faint, but rather inhomogeneous in terms of count rates in each wavelength range. The spatial distribution of NOID sources is only weakly concentrated toward the Galactic disk. In this study we concentrate on a subclass of *EUVE* NOID sources, whose flux in the longest wavelength band (Tin) is remarkably strong. Specific sources associated with this class are EUVE J0256+080, EUVE J1244-596, EUVE J1533+337, EUVE J1636-285 and EUVE J1748+484 (The 2nd *EUVE* catalog). The hydrogen column density estimated for the space positions of these sources suggest that none of them is likely to be more than 100pc away (B. Welsh, private communication). For neutral hydrogen column density $N_{HI} = 10^{17}$ cm$^{-2}$ the ISM transmission in the Tin band is $\sim 0.7$, the same order of magnitude as the transmission in the Lex and AlC bands ($\sim 1$). However, when the column density is raised to $10^{18}$ cm$^{-2}$, the Tin-to-Lex transmission ratio is lowered to $\sim 0.1$, and when $N_{HI} = 10^{19}$ cm$^{-2}$ this transmission ratio reduces to $\sim 10^{-10}$ (!). Therefore, independent of their spectral distribution, the column density towards sources of 'Tin excess' flux is strongly constrained, to $N_{HI} \lesssim 10^{18}$ cm$^{-2}$.

## 2  THE EUV COUNT RATES OF AN ACCRETING NEUTRON STAR

In hydrodynamical spherical accretion (Bondi & Hoyle 1944) the accretion rate is strongly dependent on the inverse of the star's velocity and the density of the ambient matter ($\dot{M} \propto nv^{-3}$). The average speed of Galactic neutron stars (Madau & Blaes 1994) is believed to be rather large, $\sim 100$ km s$^{-1}$. Only matter enclosed within the accretion radius $r_{acc} \sim 3.7 \times 10^{12} v_{100}^{-2}$ cm (where $v_{100} \equiv v/100$ km s$^{-1}$) could be accreted onto the neutron star. The neutral gas density in the local interstellar medium is unusually low, estimated to be $\lesssim 0.01$cm$^{-3}$ for most directions out to 50 pc (Cox & Reynolds 1987, Welsh et al. 1994). Accretion onto close neutron stars would therefore occur on low rate. Moreover, the accretion efficiency is even further lowered if the particles are not coupled, where the fluid approximation breaks down. In our case the gas is indeed collisionless, since $r_{acc}$ is orders of magnitude shorter than mean free path for Coulomb collisions ($\sim 10^{18} n_{0.01}^{-1}$ cm; $n_{0.01} \equiv n/0.01$ cm$^{-3}$). However, the particles, that are photoionized by emission from the accreting neutron star, are coupled by ambient magnetic fields and plasma instabilities (for $B = 10^{-6}$ G the Larmor radius is 3-4 orders of magnitude smaller than the accretion radius).

Low accretion rate

$$\dot{M} = 2.9 \times 10^7 n_{0.01} v_{100}^{-3} \text{ g s}^{-1} \quad (1)$$

could yield bolometric luminosity of only a small fraction of $L_\odot$,

$$L = f_{em} \dot{M} \frac{GM}{r} = f_{em} 1.8 \times 10^{27} \dot{M}_7 r_6^{-1} \text{ erg s}^{-1}. \quad (2)$$

Here $f_{em}$ is the fraction of the accretion luminosity being converted to radiation, M is the neutron star mass, taken to be 1.4M$_\odot$, $r_6$ is the radius in units of $10^6$ cm and $\dot{M}_7 \equiv \dot{M}/10^7$ g s$^{-1}$.

The accretion power can be converted to radiation through bremsstrahlung (mostly ion-electron, as $T \ll 0.51$ MeV), Compton and, in the case of strong magnetic fields, synchrotron processes. Treves & Colpi (1991), Blaes & Madau (1993), and Madau & Blaes (1994) assume the accreted matter is channeled by the dipole magnetic field toward the magnetic poles and the emission is released from optically thick plasma in the star's atmosphere above the polar caps. Nelson et al. (1994) argue that a few percent of the luminosity in that case should be nonthermal, emitted in Comptonized cyclotron photons just below 11.6$(B/10^{12}$G$)$ KeV. In an alternate, optically thin model, particles are accelerated at the Alfvén surface, far from the neutron star in a standing shock wave. The energetic particles upscatter the ambient soft photons and emit synchrotron radiation. The resulting emission spectrum is nonthermal, generically a power law (Shemi 1995).

The intrinsic spectrum of an EUV source is modified by the attenuation of the ISM and by the filtering of the telescope. Denoting $F$ the intrinsic photon spectral distribution, $TR$ the ISM transmission, and $A$ the telescope effective area, the resulting spectrum of the source would be $F(\lambda)TR(\lambda)A(\lambda)$. For a source at distance $d$ the overall count rate in a bandpass $\Delta\lambda$ is the integral over this spectrum,

$$\text{count rate} = \frac{1}{4\pi d^2} \int_{\Delta\lambda} F(\lambda) TR(\lambda) A(\lambda) d\lambda. \quad (3)$$

In this study we have calculated the count rates in the four *EUVE* bandpasses, assuming the radiation is released from optically thick regions above the neutron star magnetic polar caps. The blackbody effective temperature in that model is given by

$$T_{eff} \simeq 20 f_{em}^{1/4} \dot{M}_7^{1/4} A_{cap,1}^{-1/4} \text{ eV} \quad (4)$$

where $A_{cap,1}$ is the emitting area, in units of 1 km$^2$.

Figure 1 show spectra of optically thick sources with blackbody at $T_{eff} = 2$, 4, and 6 eV, respectively. For each temperature we have calculated several cases, corresponding to variable column density from 0 to $10^{19}$ cm$^{-2}$. The three cases in the left side of the figure are related to small H I column density $N_{HI} = 10^{17}$ cm$^{-2}$. These spectra found to be similar to spectra obtained with a vanishing $N_{HI}$, which means that the blackbody curve is mostly modified by the telescope filtering. The cases in the right side correspond to strong ISM attenuation, $N_{HI} = 10^{18}$ cm$^{-2}$. Table 1 show the wavelength-integrated count fluxes (equation 3) for the four $N_{HI}$ values we have used (numerical values in Fig 1 and Table 1 are normalized with a Tin count rate of 400 count/Ksec). The data for effective areas vs. wavelength, and models of ISM transmission vs. wavelength and hydrogen column density, have been taken from the *EUVE* cycle III NRA. The data is available for the public in the *EUVE* WWW node.

## 3  DISCUSSION

The most prominent conclusion arises from Figure 1 and Table 1 is that the 'Tin excess' NOIDs are likely to be rather cool, $T_{eff} \lesssim 6$eV and have low column density, $N < 10^{18}$ cm$^{-2}$. The distance of these sources, unless are located in an extremely rarefied 'tunnel' in the ISM, is constrained to $d \lesssim 30 \, n_{0.01}^{-1}$ pc. Moreover, as is demonstrated in the cases of $T = 6$ eV, even for negligible ISM attenuation our model excludes hot sources ($T \gg 6$ eV).

The bolometric magnitude of a source with $L = 2 \times 10^{27}$ erg s$^{-1}$ is 19.75. Using $T_{eff} = 25,000$ K$^0$ ($B - V = -0.2$) and bolometric correction $m_{bol} - m_v = 2.6$, we obtain the V (B) magnitude 22.26 (22.06) + 5 log $d_{10pc}$ ($d_{10} \equiv d/10$ pc). The Tin count rate of such a source is found to be $\sim 50 \, d_{10}^{-2}$ count/Ksec. Taking these values, we can predict the distance of EUVE J0256+080 (Tin=256 ± 57), EUVE J1244-596 (199 ± 43), EUVE J1533+337 (131 ± 34), EUVE J1636-285 (410 ± 79) and EUVE J1748+484 (133 ± 32) to be $\sim$ 35, 40, 50, 28 and 50 $\times \dot{M}_7$ pc, respectively.

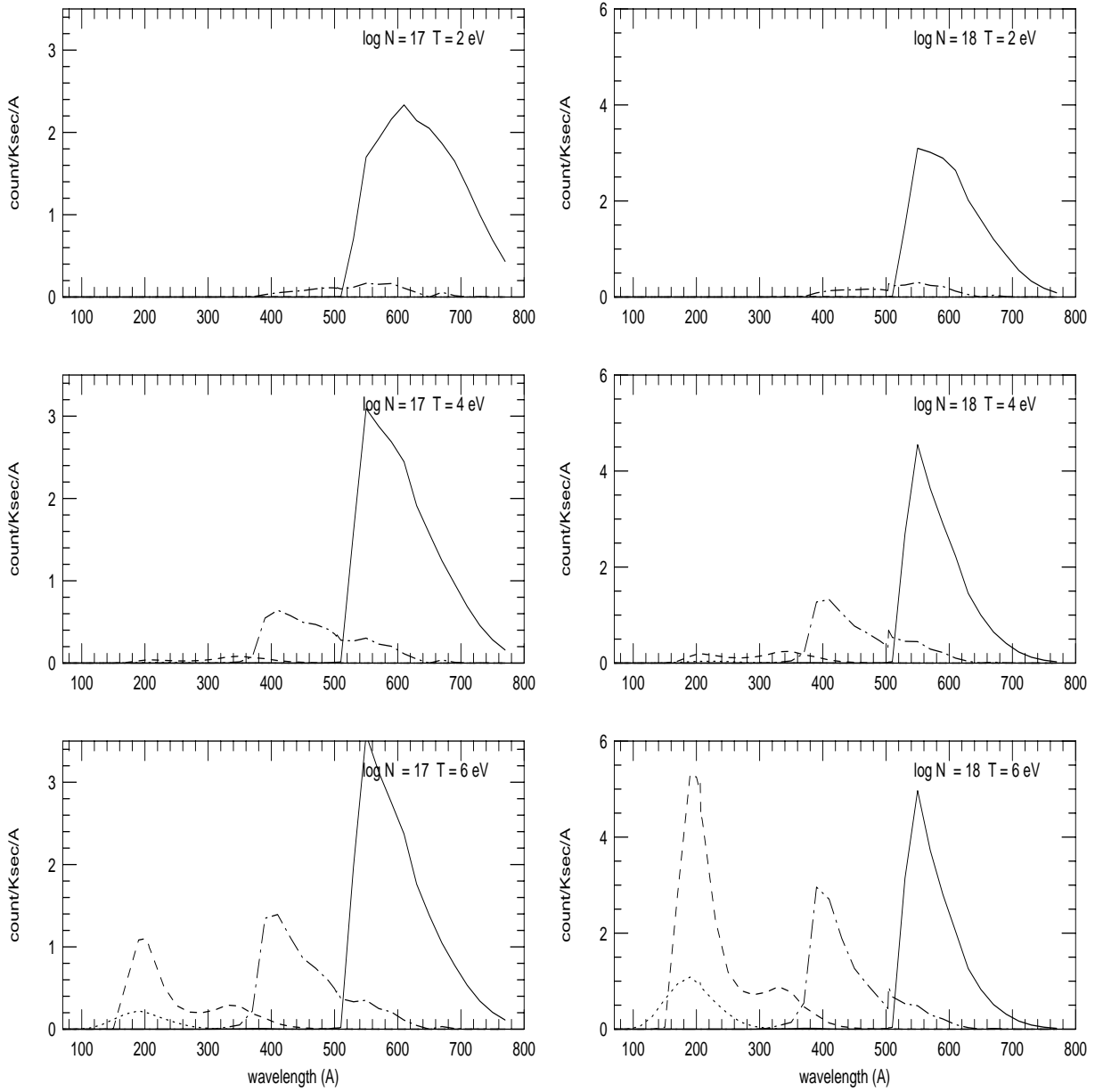

**Figure 1.** Spectra of accreting neutron stars, for $N_{HI} = 10^{17}$ (left) and $10^{18}$ cm$^{-2}$ (right). Upper: $T = 2$ eV, central: $T = 4$ eV, lower: $T = 6$ eV. Spectra corresponding to the Tin filter are plotted with solid lines. Spectra of Lex, AlC and Dag are drawn with dotted, short-dashed and dotted-dashed lines, respectively.

Table 1. Count rates in EUVE bandpasses

| $N_{HI}/10^{18}$ | Lex | AlC | Dag | Tin |
|---|---|---|---|---|
| $T_{eff} = 2$ | | | | |
| 0 | 0 | 0 | 26 | 400 |
| 0.1 | 0 | 0 | 28 | 400 |
| 1 | 0 | 1 | 45 | 400 |
| 10 | 192 | 2052 | 843 | 400 |
| $T_{eff} = 4$ | | | | |
| 0 | 1 | 10 | 90 | 400 |
| 0.1 | 1 | 12 | 96 | 400 |
| 1 | 4 | 39 | 156 | 400 |
| 10 | $4.2 \times 10^5$ | $1.7 \times 10^6$ | 4784 | 400 |
| $T_{eff} = 6$ | | | | |
| 0 | 17 | 88 | 153 | 400 |
| 0.1 | 21 | 103 | 163 | 400 |
| 1 | 101 | 431 | 268 | 400 |
| 10 | $10^7$ | $2.6 \times 10^7$ | 5889 | 400 |

So far we have taken the conservative approach and considered sources whose spectra are blackbody. Photoionization of He I and He II in the source atmosphere could, however, reduce the photon flux below the He I and He II Lyman edges (504Å and 228Å, respectively) and significantly reduce the EUV-to-UV flux ratio. Such continuum 'jumps', of about two orders of magnitude, are predicted in models of early type stars (Kurucz 1979). The spectrum of the strongest EUV source detected so far, the B2 II star $\epsilon$ CMa, indeed confirms atmospheric He I and He II attenuation, although the observed jumps are shallower than predicted (Cassinelli et al. 1995). In order to take into account such an atmosphere, whether viable or not, we included synthetic jumps in spectra that are inputed in our calculations. We found only a moderate effect in the spectrum. The ISM attenuation, once it is not negligible, is always the most important factor in modifying the blackbody curve. The conclusion is that, unlike accreting neutron stars in binary systems, hot ($T_{eff} \sim 100$ eV) sources remain improbable.

Using equation 4 one can see how the low temperature we obtained strongly constrains the physical parameters of the source, $f_{em}\dot{M}_7 A_{cap,1}^{-1} < 10^{-2}$. With this constrain, assuming large efficiency $f_{em} \sim 1$ and plausible accretion rate $\dot{M}_7 \geq 1$, the emitting area required to be much larger than in X-ray binaries. In X-ray binaries the emission is believed to be released from an accretion column, that is supported by a strong dipole magnetic field. If the emitting area in our case is hundreds time larger, the magnetic field should be significantly weaker. For $T_{eff} \lesssim 3$ eV Eq. 4 translates to an emitting area comparable with a neutron star surface, $\gtrsim 10^{13}$ cm$^2$. One can infer that the flow configuration remains spherical even near the neutron star. Magnetic field pressure is insufficiently strong to balance ram pressure of the flow when approaching the star. When the flow is converged the ram pressure is enhanced, roughly by $(r_{acc}/r_6)^{5/2} \sim 3 \times 10^{16}$. From the pressure imbalance we find an upper limit on the magnetic field $B \leq 750 n_{0.01}^{1/2} v_{100}^{-3/2}$ G. Such a field is ten orders of magnitude (!) lower than magnetic fields in pulsars. Evidently, we consider this value with great caution and leave it to be studied elsewhere, as the detailed physics of the accretion stands beyond the scope of this *Letter*. Nevertheless, it is worth remarking that such a low magnetic field, if real, could be evidence for strong field decay.

If they are not neutron stars, what could the EUV 'Tin excess' NOIDs be? Assume the source of the EUV radiation is an early type star like $\epsilon$ CMa, with similar count rate ratios and optical luminosity ($\epsilon$ CMa : Lex : AlC : Dag : Tin = 0 : 80 : 4030 : 41840, $m_v = 1.5$, $T_{eff} \sim 21,000$ K$^0$). Simple scaling gives a bright $m_v \sim 6.5$ star. As no optical counterparts brighter than $m_v = 14$ appear within the error circle of the NOID sources mentioned above, early type stars are unlikely. Other EUV sources like white dwarfs or late type stars, if having $M_v \lesssim 14$, are also improbable, while extensive optical observations of the NOID fields are required to constrain fainter optical counterparts (Maoz et al. 1995).

Selection and a comprehensive study of best neutron star candidates among the NOID sources need deep optical UV, EUV and X-ray observations. Simultaneous observations, for instance with the future Russian - International *Spectrum-X-Gamma* observatory, will be very useful to constrain eruptive sources, like flare stars or cataclysmic variables. Moreover, repeat of observations with good angular resolution, on time scale of a few years, could confirm proper motions of $2'' \, d_{10}^{-1} v_{100} \sin\theta$ yr$^{-1}$, where $\theta$ is the angle between the line of sight and the star trajectory.

The detection of isolated old neutron stars will provide extremely valuable data on the neutron star physics and evolution. Since isolated neutron stars are the fossils of the high-mass end of the initial mass function, their detection could provide important clues for deciphering the stellar evolution history of the Galaxy. Isolated neutron stars may also be important as Galactic gamma-ray burst sources (Higdon & Lingenfelter 1990), cosmic ray accelerators (Shemi 1995), and the origin of the Galactic excess in the X-ray background (Maoz & Grindlay 1994).

### Acknowledgments

Astronomy at the Wise Observatory is supported by grants from the Ministry of Science and the Arts and from the Israel Academy of Science. The author acknowledges discussions with Piero Madau, Dan Maoz and Barry Welsh.